\documentclass[12pt,preprint]{aastex}
\RequirePackage{color}

\shorttitle{Near-IR Polarimetry toward Serpens South}
\shortauthors{Sugitani et al.}

\begin{document}

\title{Near-Infrared Imaging Polarimetry Toward Serpens South: 
Revealing the Importance of the Magnetic Field
}
\author{K. Sugitani,\altaffilmark{1} 
F. Nakamura,\altaffilmark{2}   
M. Watanabe,\altaffilmark{3} 
M. Tamura,\altaffilmark{2} 
S. Nishiyama,\altaffilmark{4} 
T. Nagayama,\altaffilmark{5} 
R. Kandori,\altaffilmark{2} 
T. Nagata, \altaffilmark{4} 
S. Sato,\altaffilmark{5} 
R. A. Gutermuth, \altaffilmark{6} 
G. W. Wilson,\altaffilmark{7} 
and 
R. Kawabe \altaffilmark{8} 
}
\altaffiltext{1}{Graduate School of Natural Sciences, 
Nagoya City University, Mizuho-ku, Nagoya 467-8501, Japan; 
sugitani@nsc.nagoya-cu.ac.jp.}
\altaffiltext{2}{National Astronomical Observatory, Mitaka, Tokyo 181-8588, Japan.} 
\altaffiltext{3}{Department of Cosmosciences, Hokkaido University, Kita-ku, 
Sapporo, Hokkaido 060-0810, Japan.}
\altaffiltext{4}{Department of Astronomy, Kyoto University, Sakyo-ku, Kyoto 606-8502, 
Japan.}
\altaffiltext{5}{Department of Astrophysics, Nagoya University, Chikusa-ku, 
Nagoya 464-8602, Japan.}
\altaffiltext{6}{Smith College, University of Massachusetts, Northampton, MA 01063.}
\altaffiltext{7}{Department of Astronomy, University of Massachusetts, Amherst, MA 01003.}
\altaffiltext{8}{Nobeyama Radio Observatory, Nobeyama, Minamimaki, Minamisaku, 
Nagano 384-1305, Japan.}

%%%% limit of number of words is 250.
\begin{abstract}
The Serpens South embedded cluster, which is located at the constricted part in a long 
filamentary infrared dark cloud,  is believed to be in very early stage of cluster formation.   
We present results of near-infrared ({\it JHK}s) polarization observations 
toward the filamentary cloud.
Our polarization measurements of near-infrared point sources indicate a well-ordered 
global magnetic field that is perpendicular to the main filament, implying that 
the magnetic field is likely to have controlled the formation of the main filament.  
On the other hand, the sub-filaments, which converge on the central part of the cluster, 
tend to run along the magnetic field.
The global magnetic field appears to be curved in the southern part of the main filament. 
Such morphology is consistent with the idea that the global magnetic field is distorted 
by gravitational contraction along the main filament toward the northern part that contains 
largermass.
Applying the Chandrasekhar-Fermi method, the magnetic field strength is roughly estimated to be 
a few $\times$100 $\mu$G, suggesting that the filamentary cloud is close to magnetically critical 
as a whole.
\end{abstract}

\section{Introduction}
\label{sec:intro}

It is now widely accepted that stars form predominantly within 
clusters inside dense clumps of molecular clouds
that are turbulent and magnetized.
However, how clusters form in such dense clumps remains poorly understood.
This is due in part to the lack of the observational characterization
of processes involved in cluster formation.
In particular, the role of magnetic fields in cluster formation
is a matter of debate.

Recent numerical simulations of cluster formation suggest that 
a moderately-strong magnetic field is needed to impede star formation 
in molecular clouds in order for the simulated star formation rates to
match observed values 
\citep{enrique05,li06, nakamura07, price08}.
In contrast, \citet{padoan01} claim that 
the magnetic field in molecular clouds should be significantly 
weak (a few $\mu$G) and that the turbulent compression 
largely controls the structure formation in molecular clouds 
on scales of a few to several parsecs.
Such a very weak magnetic field is necessary to reproduce 
the core mass spectrum that resembles the stellar 
initial mass spectrum in shape, in the context of their turbulent 
fragmentation scenario \citep{padoan02}.
In this picture, strong magnetic fields sometimes observed in 
dense clumps and cores are due to local amplification by the 
turbulent compression.  Recent Zeeman measurements in molecular clouds 
and cores appear to support this idea \citep{crutcher10}.
If this is the case, the magnetic fields associated with cluster-forming
clumps are expected to be distorted significantly 
by the supersonic turbulent flows.

In order to characterize the magnetic structure of cluster-forming
clumps, it is important to uncover the global magnetic field structures
associated with as many cluster forming clumps as possible.
Polarimetry of background star light is one of the techniques suitable 
for mapping the magnetic fields in molecular clouds and cores 
\citep{tamura87} and the technique is now strengthened with wide-field 
polarimetry by employing large array detectors \citep[e.g.,][]{tamura07}.
As a continuation of our recent polarimetry study of the Serpens cloud core 
\citep{sugitani10}, which revealed an hour-glass shaped magnetic field 
that implies the global gravitational contraction of the cluster forming clump, 
we have made near-infrared polarization observations 
toward Serpens South. 

Serpens South is a nearby ($d\sim$260 pc), embedded cluster, 
discovered by \citet{gutermuth08} in the context of 
the Spitzer Gould Belt Legacy Survey.
The Serpens South cluster appears to be located at the constricted region
in a long filamentary cloud or at the joint region of multiple 
less dense filaments - now referred to as a "hub-filament" structure \citep{myers09}.
The Serpens cloud core resembles such morphology, although the filamentary 
structures are much more prominent around Serpens South. 
In the central part of the cluster, the fraction of protostars (Class 0/I sources) 
reachs 77 \% with a high surface density of 430 pc$^{-2}$.
This fraction is largest among the cluster-forming regions within the nearest 400 pc.
Recently, \citet{bontemps10} discovered 5 Class 0 sources 
in the central core as a part of the Herschel Gould Belt Survey.
These observations  strongly suggest a very recent initialization 
of star formation in this region.
Therefore, Serpens South is one of the best objects to characterize
the initial conditions of cluster formation.

Here, we present the results of our near-infrared 
polarization measurements, and discuss the role of magnetic fields 
in the formation of this region.

\section{Observations and Data Reduction}
\label{sec:obs}

Simultaneous polarimetric observations were carried out 
toward Serpens South in $JHK$s-bands 
on 2009 August 28 and September 3, 
and 2010 June 25, August  9, 12, 13, 14, and 15  UT 
with the imaging polarimeter SIRPOL 
\citep[polarimetry mode of the SIRIUS camera:][]{kandori06} 
mounted on the IRSF 1.4 m telescope at the South Africa
Astronomical Observatory. 
The SIRIUS camera is equipped with three 1024 $\times$ 
1024 HgCdTe (HAWAII) arrays, $JHK$s filters, and dichroic mirrors, 
which enables simultaneous $JHK$s observations 
\citep{nagashima99,nagayama03}.
The field of view at each band is $7'.7 \times 7'.7$ 
with a pixel scale of 0''.45, and a 3$\times$3 mosaic area 
centered around (R.A., decl.)$_{\rm J2000}$=(18$^h$30$^m$05$^s$, -02\degr03')
was covered, including duplicate field observations. 

We obtained 10 dithered exposures, each 10 or 15 s long, at four wave-plate 
angles (0$^\circ$, 22$^\circ$.5, 45$^\circ$, and 67$^\circ$.5 in the instrumental
coordinate system) as one set of observations and repeated this
nine or six times, respectively. 
Thus, the total on-target exposure time for each field was 900 s 
per wave-plate angle. 
Sky images were also obtained in between target observations.
Self sky images were also used with the sky images 
to make better median sky images.
The average seeing was $\sim$1.$\arcsec$4 at $K$s during the
observations with a range of $\sim$1.$\arcsec$2--1.$\arcsec$6. 
Twilight flat-field images were obtained at the
beginning and/or end of the observations.
Standard reduction procedures were applied with IRAF. 
Aperture polarimetry was performed at $H$ and $K$s 
with an aperture of $\sim$FWHM
by using APHOT of the DAOPHOT package. 
No polarimetry for J band sources was done due to their much smaller number, 
compared with those of $H$ and $K$s band sources.  
The 2MASS catalog \citep{skrutskie06} was used for absolute photometric calibration.
See \citet{sugitani10} for more details of the data reduction procedure and 
the method to derive the polarization degree ($P$) and its error ($\Delta P$), 
and the polarization angle ($\theta$ in P.A.) and its error ($\Delta \theta$).

\section{RESULTS and DISCUSSION}
\subsection{Magnetic Field Structure toward Serpens South}

We have measured $H$ and $K$s polarization for point sources, 
in order to examine the magnetic field structure.
Only the sources with photometric errors of $<$0.1 mag and 
$P/{\mathit \Delta} P > 3.0$ were used for analysis. 

\begin{figure}[h]
\plotone{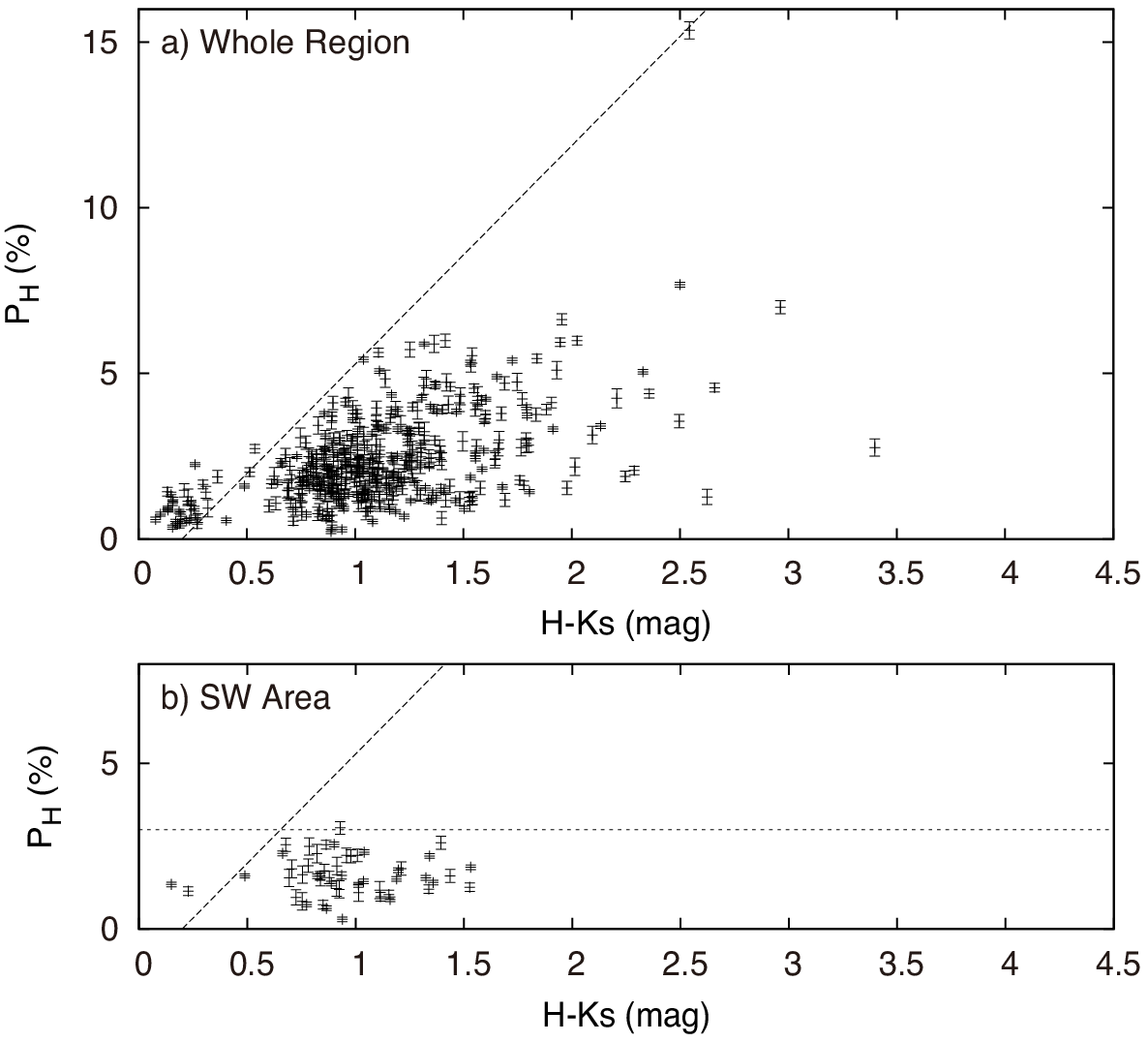}
\caption{Polarization degree at $H$ versus 
$H-K$s color diagram for sources having polarization errors of $<0.3\%$ 
in (a) the whole region and (b) the SW area. 
YSOs/YSO candidates identified by \cite{gutermuth08} and \cite{bontemps10} 
are not included. 
The lines of the adopted maximum polarization efficiency of 
$P_{H} = 6.6([H-K{\rm s}]-0.2)$ are shown both in the top and bottom 
panels, and the line of $P_{H}=3.0$ is shown in the bottom panel. 
}  
\label{f1}
\end{figure}

\begin{figure}[h]
\plotone{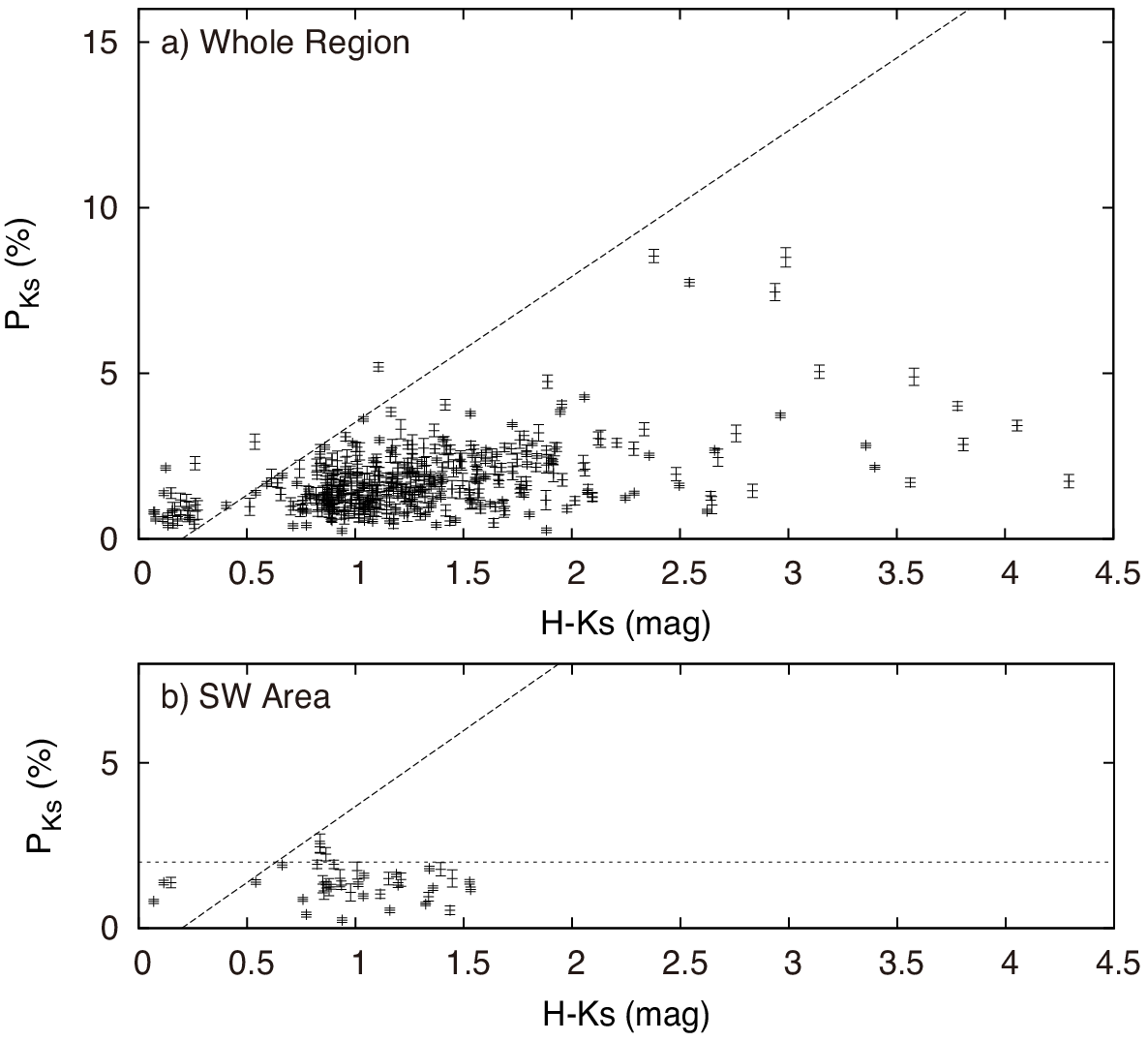}
\caption{Polarization degree at $K$s versus 
$H-K$s color diagram for sources having polarization errors of $<0.3\%$ 
in (a) the whole region and (b) the SW area. 
YSOs/YSO candidates identified by \cite{gutermuth08} and \cite{bontemps10} 
are not included. 
The lines of the adopted maximum polarization efficiency of 
$P_{K{\rm s}} = 4.4([H-Ks]-0.2)$ are shown both in the top and bottom 
panels, and the line of $P_{K{\rm s}} =2.0$ is shown in the bottom panel. 
}  
\label{f2}
\end{figure}

Figures \ref{f1}a and \ref{f2}a present the polarization degree versus 
the $H-K$s color for sources having polarization errors of $<0.3\%$ at $H$ and $K$s, 
respectively.   
There is a tendency for the upper envelope of the plotted points to increase with $H-K$s, 
and the average polarization degree is slightly smaller at $K$s than at $H$ 
for the same $H-K$s color.
These are consistent with the origin of the polarization being dichroic absorption.  
Therefore, here we assume the polarization vectors as the directions of 
the local magnetic field average over the line of sight of the sources.
These sources, except those of low $H-K$s colors, appear to have the maximum 
polarization efficiencies of $P_H/([H-K\rm{s}]-0.2) =6.6$ and 
$P_{K_{\rm s}}/([H-K\rm{s}]-0.2)=4.4$, 
where we adopt $H-K\rm{s}=0.2$ as an offset of the intrinsic $H-K\rm{s}$ color, 
because the 2MASS point sources with good photometric qualities 
are mostly located on the reddening belt that begins from the position of 
($J-H$, $H-K$s)=($\sim$0.7, $\sim$0.2) on the $J-H$ versus $H-K$s diagram, 
within a $1\degr \times 1\degr$ area centered at Serpens South. 

\begin{figure}[h]
\plotone{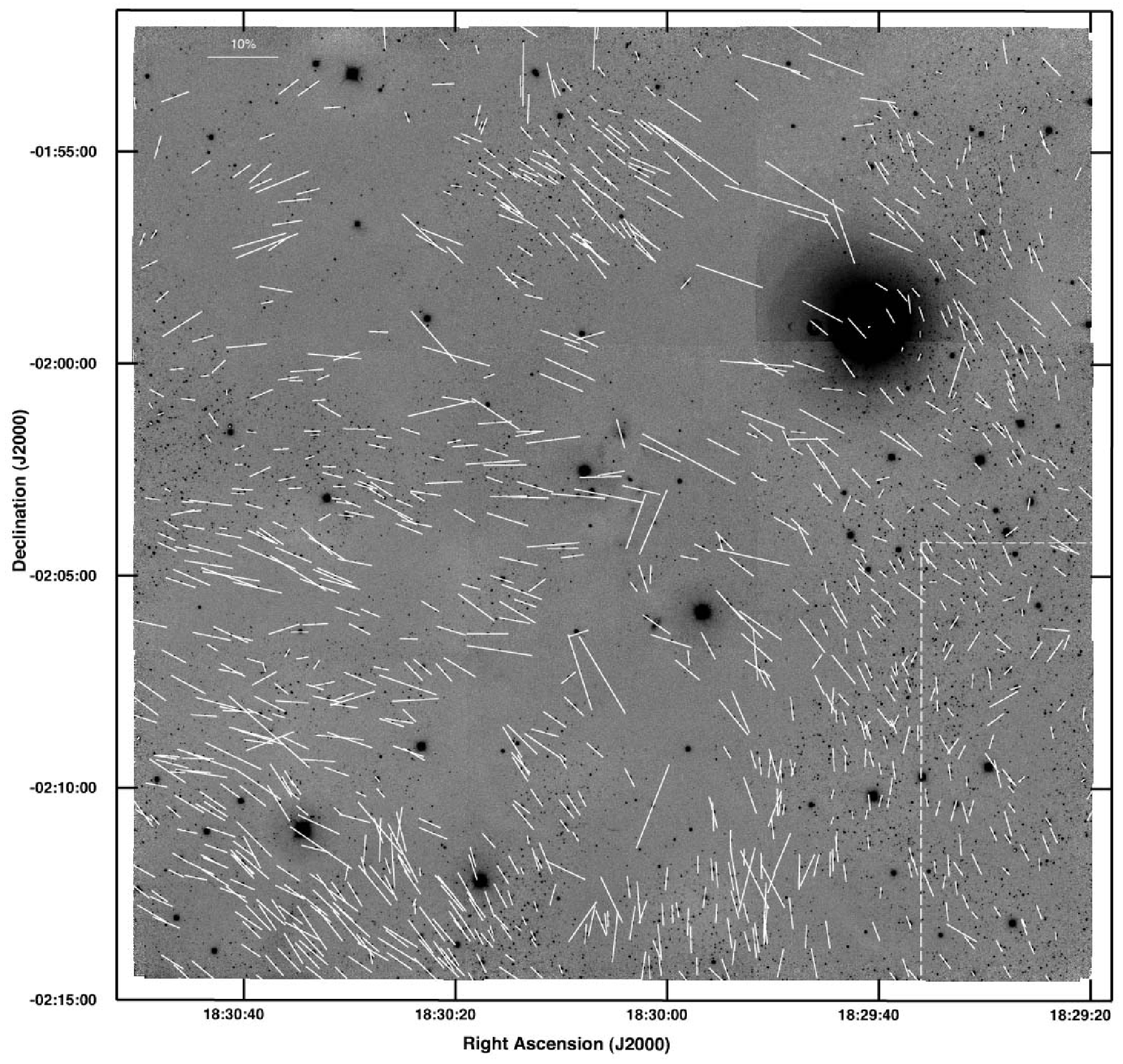}
\caption{$H$-band polarization vector map toward Serpens South for point sources 
having $P/{\mathit \Delta} P > 3.0$ and $P < 6.6([H-Ks]-0.2)$, 
superposed on the $H$-band image. 
The Serpens South cluster is located toward the center of the image.  
}  
\label{f3}
\end{figure}

\begin{figure}[h]
\plotone{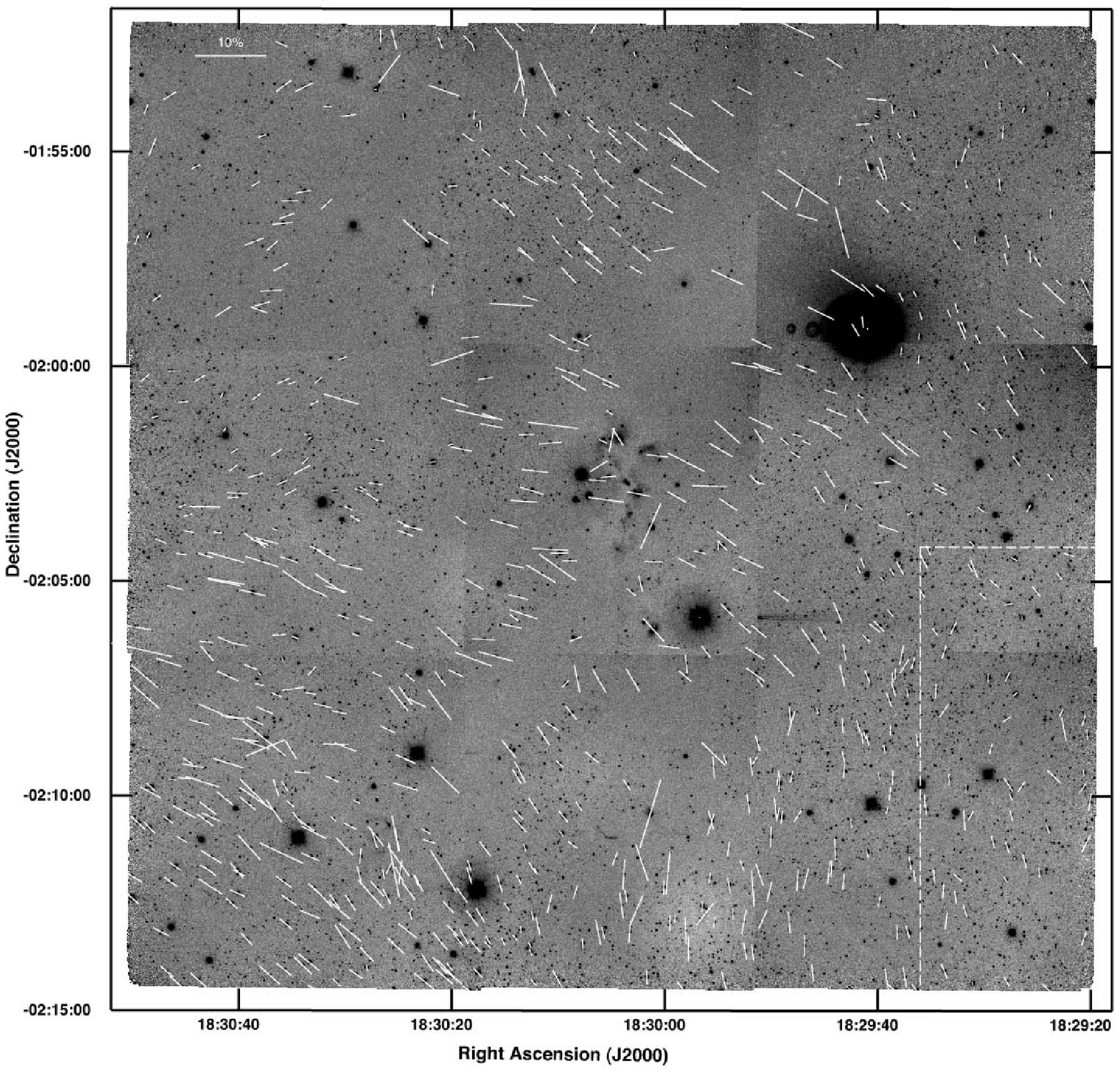}
\caption{$K$s-band polarization vector map toward Serpens South for point sources 
having $P/{\mathit \Delta} P > 3.0$ and $P < 4.4([H-Ks]-0.2)$, 
superposed on the $K$s-band image. 
The Serpens South cluster is located toward the center of the image.  
}  
\label{f4}
\end{figure}

\begin{figure}[h]
\plotone{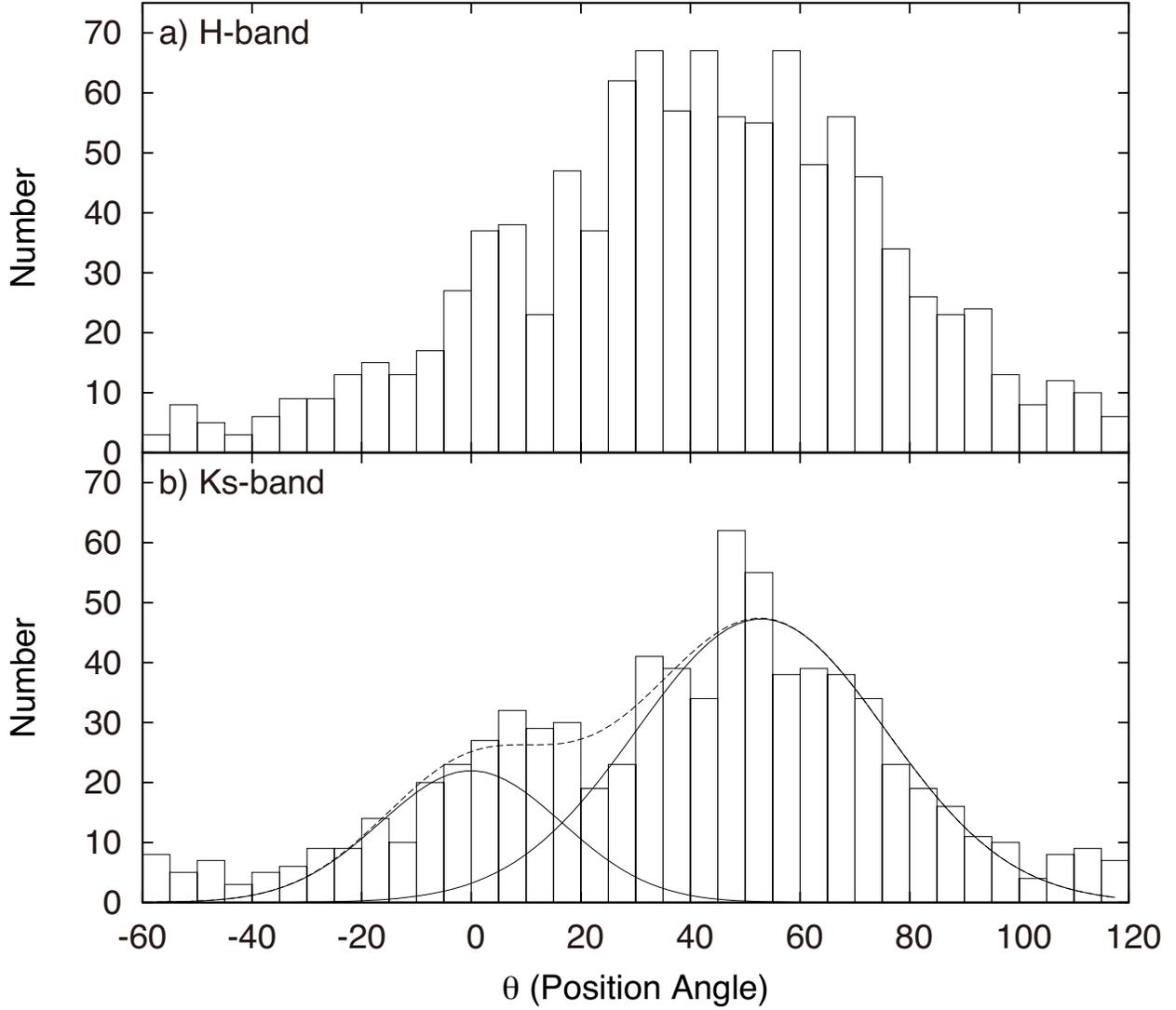}
\caption{Histograms of the position angles of  
a) $H$-band vectors for point sources of $P/{\mathit \Delta} P > 3.0$ and $P < 6.6([H-Ks]-0.2)$, 
and b)  $K$s-band vectors for point sources of $P/{\mathit \Delta} P > 3.0$ and $P < 4.4([H-Ks]-0.2)$.
A two-Gaussian fit is shown on the $K$s-band histogram. }  
\label{f5}
\end{figure}

Figures \ref{f3} and \ref{f4} present $H$ and $K$s-band polarization vectors of 
$P/{\mathit \Delta} P > 3.0$, 
excluding sources with polarizations larger than the above maximum polarization, 
superposed on 3$\times$3 mosaic $H$ and $K$s-band images, respectively.
In general, the global magnetic field structures deduced from the $H$ and 
$K$s-band data seem to be the same.
Most of the vectors are roughly aligned with the NE-SW 
direction with the exception of those appearing in the NE and SW corners of the map.
Two distribution peaks are clearly seen in the histogram of the $K$s-band vector 
angles, although a sub-peak at $H$ can be barely seen (Figure \ref{f5}).
Our two-Gaussian fit analysis for the $K$s-band vectors indicates that 
the mean angle of the main peak is $52.9\degr \pm3.0\degr$ with a standard deviation of 
$22.8\degr\pm2.8\degr$ and that the mean angle of the sub-peak is $0.0\degr\pm5.2\degr$ 
with a standard deviation of $16.6\degr\pm4.3\degr$.
The low number density of background stars toward the NE corner area 
suggests a molecular cloud or filament, other than the Serpens South cloud, 
having a different magnetic field structure.
On the other hand, toward the SW corner area, no clear signs of clouds 
are seen with high density of background stars.
To investigate this further we examined the sources within
a zone enclosed by the dotted lines toward the SW corner of the maps.
Unlike in the main filament, there is no tendency that the degree of polarization 
in this zone increases as an increase of $H-K$s color, and the degrees of polarization are 
relatively low with values $\lesssim$3\% at $H$ and $\lesssim$2\% at $K$s 
(except a few sources) in Figures \ref{f1}b and \ref{f2}b.  
Thus, the polarization in this SW area may not be dominated by the dichroic absorption 
of the Serpens South cloud and is likely contaminated by either foreground or
background interstellar polarization, 
although it could be also possible that the-line-of-sight component 
of the magnetic field is dominant toward the SW corner area.

\begin{figure}[h]
\plotone{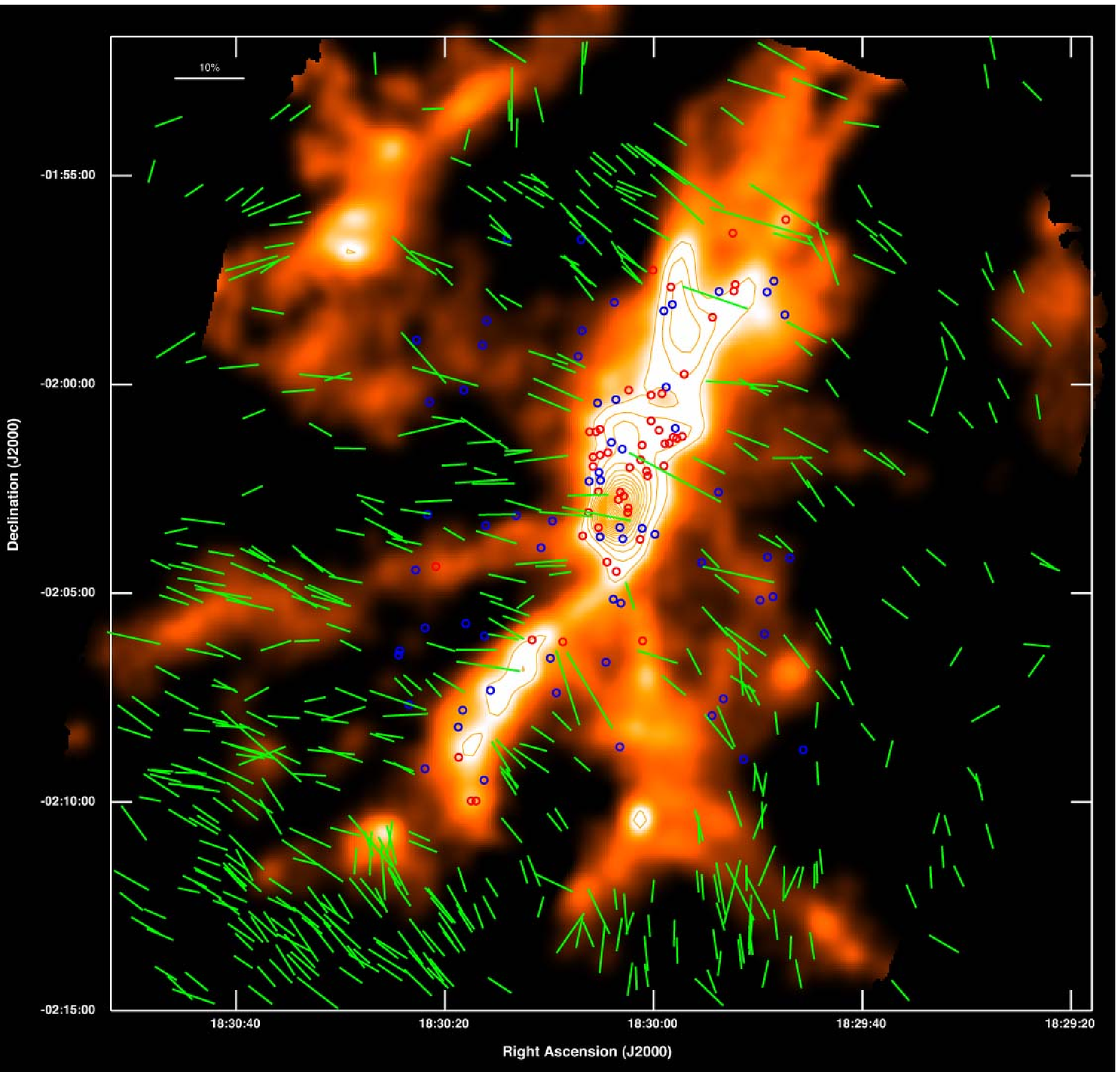}
\caption{$H$-band polarization vector map toward Serpens South for point sources 
having $P/{\mathit \Delta} P > 3.0$, $P < 6.6([H-Ks]-0.2)$, and $P>3.0$\%,
superposed on 1.1mm dust continuum image of ASTE/AzTEC  \citep{gutermuth11}. 
YSOs identified by \cite{gutermuth08} and  \cite{bontemps10} are not included,   
but those of \cite{gutermuth08} are indicated by red (class 0/I) and blue (class II) open circles.
}  
\label{f6}
\end{figure}

\begin{figure}[h]
\plotone{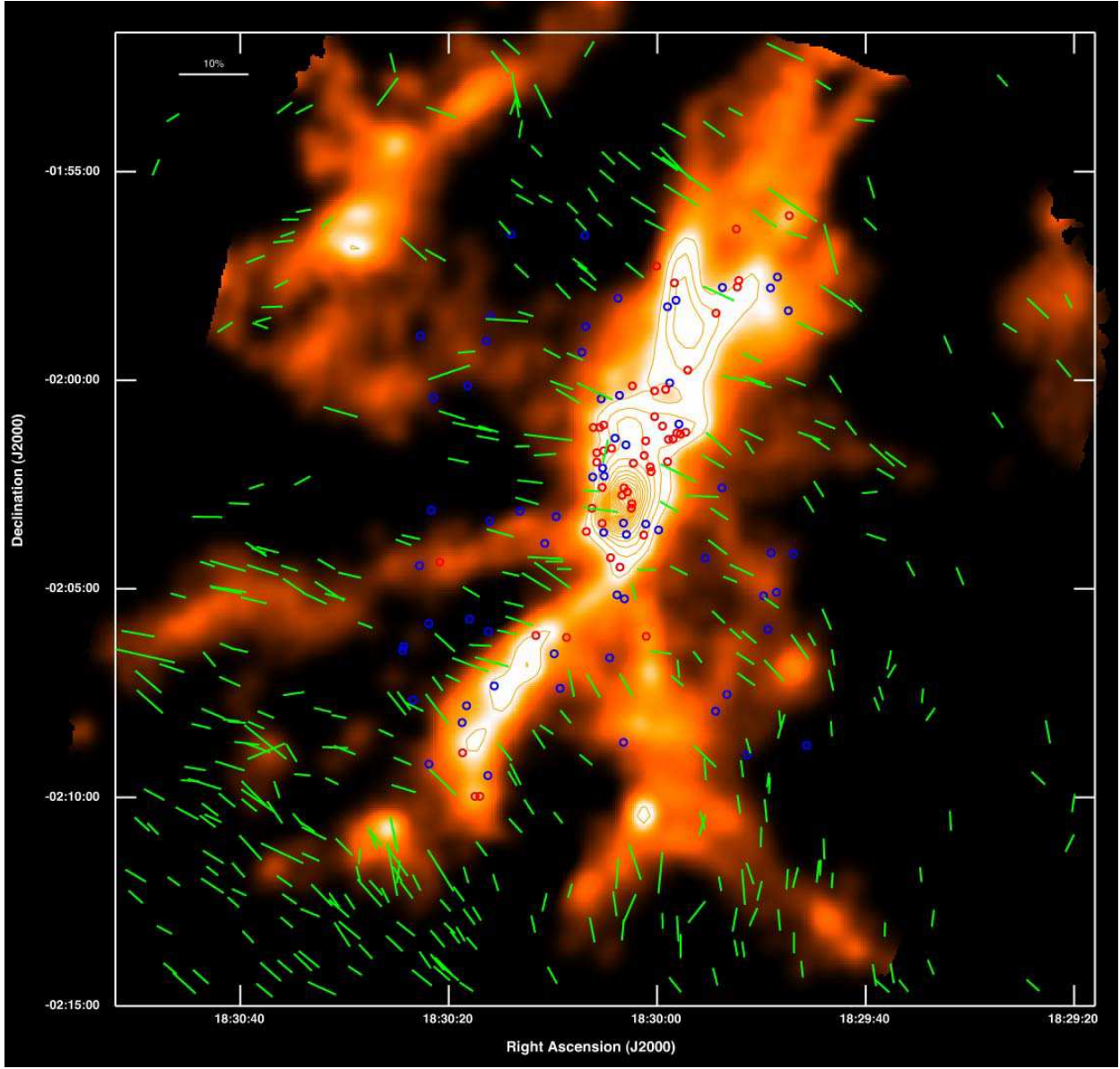}
\caption{$K$s-band polarization vector map toward Serpens South for point sources 
having $P/{\mathit \Delta} P > 3.0$, $P < 4.4([H-Ks]-0.2)$, and $P>2.0$\%,
superposed on 1.1mm dust continuum image of ASTE/AzTEC  \citep{gutermuth11}. 
YSOs identified by \cite{gutermuth08} and  \cite{bontemps10} are not included, 
but those of \cite{gutermuth08} are indicated by red (class 0/I) and blue (class II) open circles.
}  
\label{f7}
\end{figure}

\begin{figure}[h]
\plotone{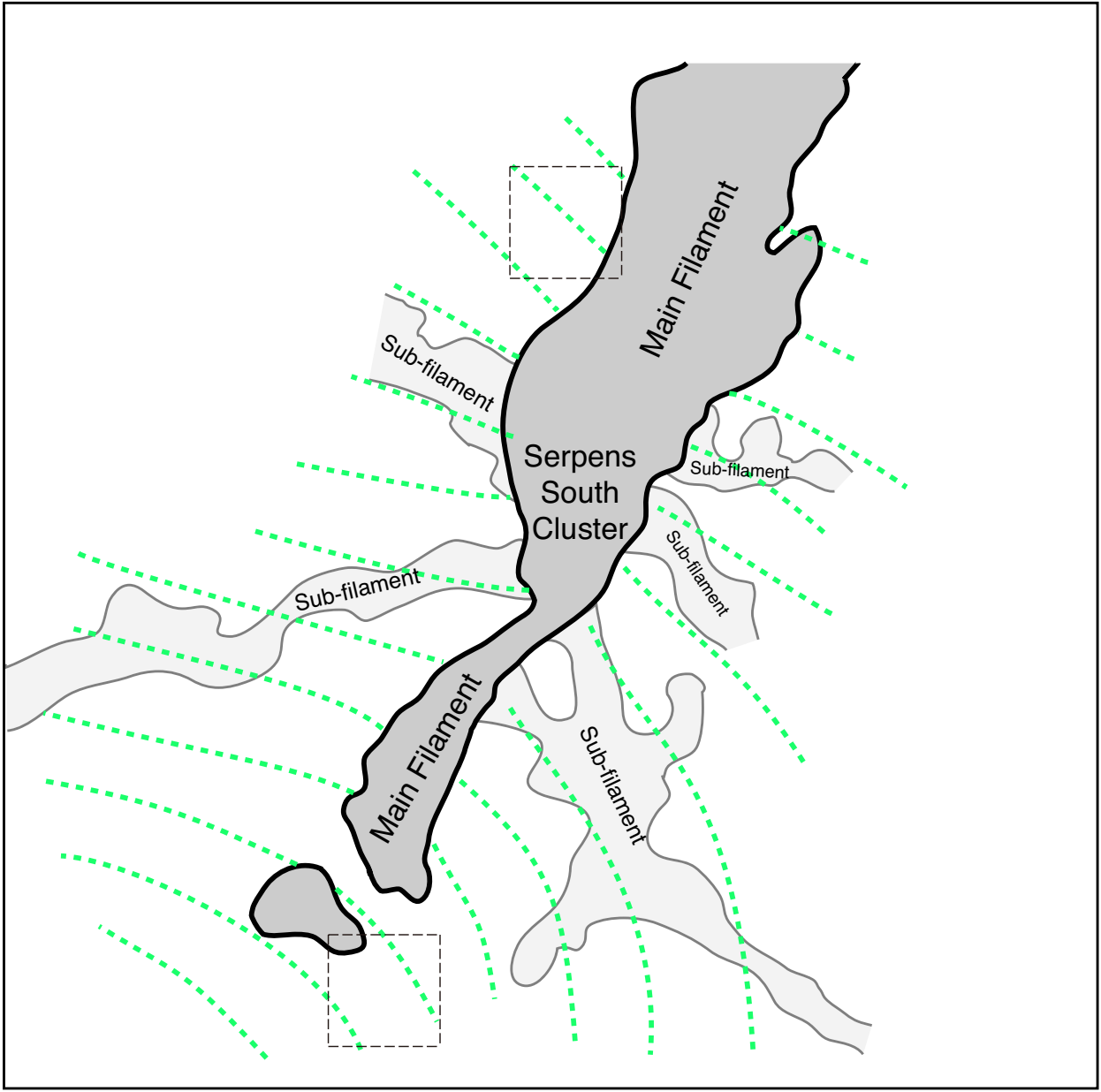}
\caption{Schematic drawing of the main and sub-filaments of the Serpens South cloud. 
The outlines of the filaments and the magnetic field lines
are shown by black lines and green dotted lines, respectively. 
The outlines of the filaments were drawn based on the 1.1mm dust continuum image 
\citep{gutermuth11} and the magnetic field lines were deduced from the $H$-band 
polarization vectors of Figure \ref{f6}. 
In this figure, the magnetic field lines presents only the direction of the magnetic field, 
not the strength of the magnetic field.
The boxes outlines by black, dotted lines indicate the zones where the strength 
of the magnetic field is roughly estimated.}  
\label{f8}
\end{figure}

We apply a cut on the degree of polarization of at least 3\% in $P_H$ 
and 2\% in $P_{K{\rm s}}$ in order to insure that we are sampling the magnetic field 
associated with the dichroic absorption of the material in Serpens South.
Figures \ref{f6} and \ref{f7} present the polarization vectors 
for the sources of $P_H>3.0$\% and $P_{K{\rm s}}>2.0$\%, resepctively,  
superposed on 1.1mm dust continuum image  \citep{gutermuth11}, 
which was taken with  the 144 pixel bolometer camera AzTEC \citep{wilson08} 
mounted on the Atacama Submillimeter Telescope Experiment (ASTE).
Figure \ref{f8} presents the schematic drawing of the filaments 
and the magnetic field directions toward the Serpens South region.  
The illustrations of the filaments were deduced from this continuum image.  
The directions of the magnetic field were deduced from the the $H$-band 
polarization vectors in Figure \ref{f6}.

The 1.1 mm dust continuum image clearly shows a main filament that is elongated  
toward the NW and SE directions from the central part of Serpens South. 
Along this main filament, the magnetic field is roughly perpendicular to it, 
although some small deviation is seen. 
This ordered magnetic configuration suggests that the magnetic field is 
strong enough to regulate the whole structure of the main filament and, 
therefore, that the formation of the main filament has proceeded 
under the influence of the magnetic field.  

\begin{figure}[h]
\plotone{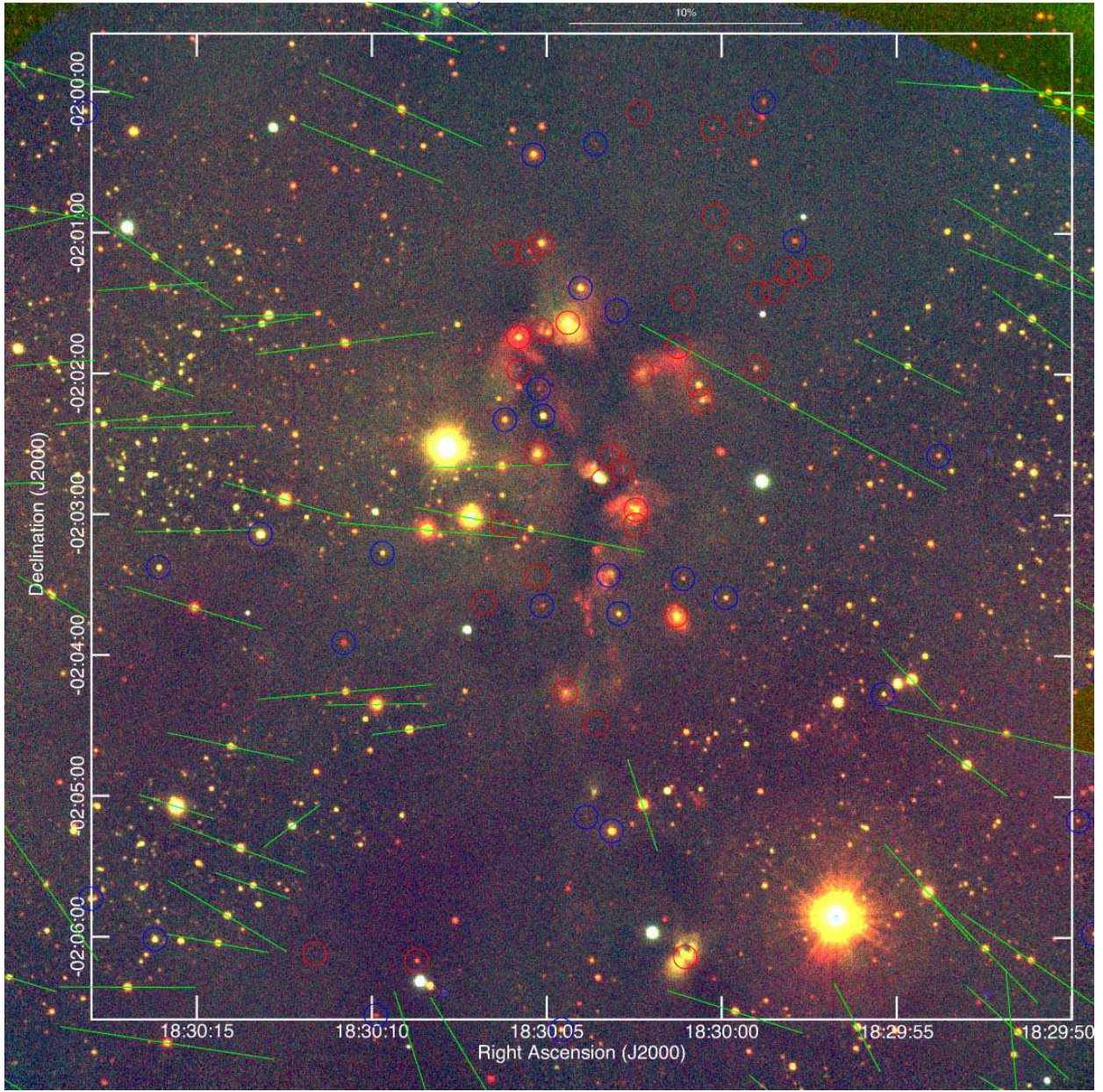}
\caption{$H$-band polarization vector map, 
superposed on a closeup, $JHK$s composite color image 
of Serpens South (R: $K$s, G: $H$, B: $J$). 
Only the vectors of the sources of $P/\Delta P >3.0$, $P<6.6([H-K{\rm s}]-0.2)$ 
and $P>3.0$\%, are shown.
YSOs identified by \cite{gutermuth08} are indicated 
by red (Class 0/I) and blue (Class II) open circles, 
but their polarization vectors are not shown.
}  
\label{f9}
\end{figure}

Also the 1.1 mm dust continuum image shows sub-filaments 
that converge on the central part of the cluster or intersect the main filament 
(Figures \ref{f6},  \ref{f7} and \ref{f8}). 
These sub-filaments are also seen in the column density map of Aquila \citep{andre10}.
In contrast to the main filament, the magnetic field appears to be nearly parallel 
to the elongation directions of the sub-filaments, 
except in some parts of the sub-filaments.   
The southern sub-filament has a more complicated structure than a simple elongated 
structure and its global elongation seems parallel to the magnetic field, 
although some parts seem perpendicular or diagonal  to the magnetic field.  
The east-southeast sub-filament is a long filament that stretches from 
the central part of the cluster toward the east-southeast direction, 
and appears to have some parts parallel to the magnetic field 
and some other parts diagonal to the magnetic field.
Near the convergent point on the main filament, 
this ESE sub-filament appears to change its elongation direction 
from the ESE-WNW direction to the E-W direction and to split into 
a few, thinner filaments that are connected to the main filament 
\citep[see Figure \ref{f9}, and also Fig. 1 of][]{gutermuth08}.
Toward this convergent point, the magnetic field also seems to be 
nearly perpendicular to the main filament just like the other sub-filaments.
These suggest that all the sub-filaments intersect the main filament 
along the magnetic field, i.e. these sub-filaments  could be 
outflows from the cluster or inflows toward the main filament.
Recent CO (3-2) observations toward Serpens South suggest that 
CO (3-2) outflow lobes are anti-correlated with the sub-filaments \citep{nakamura11}, 
preferring the inflow view of the sub-filaments.
In case of the inflows, these sub-filaments could be important 
as mass supply agents of the cluster.

Looking at the overall magnetic field structure 
in the entire observed region, we can recognize that 
the magnetic field is curved toward the cluster,
particularly in the southern area of the observed region 
(Figures \ref{f6}, \ref{f7} and \ref{f8}).
Although the origin of this large-scale curved magnetic field remains unclear,  
such morphology is consistent with the idea that 
the global magnetic field is distorted by gravitational contraction 
along the main filament toward the northern part of the main filament, 
which probably contains the majority of the mass in the Serpens South cloud.
However, we should wait for the detailed analysis of the dust continuum data 
\citep[e.g.,][]{gutermuth11} and/or molecular line data 
in order to know whether the northern part have enough mass 
to cause the large-scale curved magnetic field observed here. 

\subsection{Rough Estimate of the Magnetic Field Strength} 

We roughly estimate the magnetic field strength toward two 
(north, and south) zones enclosed by dotted lines in Figure \ref{f8}, 
where in the $H$-band polarization map (Figure \ref{f6}) 
the local number density of the polarization vectors is relatively large and 
the polarization vectors seem to be locally well-ordered, 
using the Chandrasekhar-Fermi (CF) method \citep{chadra53}.  
Here, we calculate the plan-of-the-sky component of the magnetic field strength, 
$B_\parallel$, using the equation of the CF method \citep[e.g., eq. 4 of ][]{houde04} 
and a correction factor, $C$, for the CF method \citep{houde04, houde09}, 
where we adopt $C\sim0.3$ following \cite{sugitani10}.
In this calculation, we use the $H$-band sources in Figure \ref{f6}, 
due to the sample number larger than that of the $K$s-band sources in Figure \ref{f7}.

For 21 sources toward the north zone, an average $\theta$ in P.A. is 
calculated to be 51.1$\degr\pm$9.6$\degr$, 
and an average $H-K$s color to be 1.09$\pm$0.15 mag. 
Removing the dispersion due to the measurement uncertainty of 
the polarization angle of  4.7$\degr$, the intrinsic dispersion of 
the polarization angle  ($\sigma_\theta$) of 8.4$\degr$ is obtained. 
From the average $H-K$s color, we estimate $A_{\rm V}$ by adopting 
the color excess equation of  $H-K$ color, 
$E (H-K)=[H-K]_{\rm observed} - [H-K]_{\rm instrinsic}$ \citep[e.g.,][]{lada94}, 
and the reddening law, $E(H-K)=0.065\times A_{\rm V}$ \citep{cohen81}. 
With the standard gas-to-extinction, $N_{\rm H_2}/A_{\rm V}\sim1.25\times10^{21}$ 
cm$^{-2}$ mag$^{-1}$ \citep{dickman78}, H$_{2}$ column density 
can be roughly obtained as 
$N\sim1.9\times 10^{22}\times E (H-K)$ cm$^{-2}$.
Adopting a distance from the filament axis of $\sim$2' ($l/2 \sim$0.15 pc 
at $d\sim260$ pc) as a half of the depth of this area, and 
$H-K=0.2$ as the intrinsic $H-K$ color, 
we can approximately derive a column density of $\sim1.7\times10^{22}$ cm$^{-2}$ 
and a number density ($n\sim N/l$) of $\sim1.5\times 10^4$ cm$^{-3}$.
On the basis of HCO$^+$ (4-3) observations, 
we estimate the typical FWHM velocity width of $\sim$1.5--2 km s$^{-1}$
near the cluster \citep{nakamura11}. 
Adopting this value to derive the velocity dispersion ($\sigma_{v}$), 
a mean molecular mass ($\mu$=2.3), 
and  the mass of a hydrogen atom ($m_{\rm H}$), 
we obtain $B_\parallel \sim$150 $\mu$G.
For 25 sources toward the south zone, 
with $\overline{\theta}=33.8\degr \pm 6.9 \degr$, 
$\overline{[H-K_{\rm s}]}=1.05 \pm 0.18$ mag., 
the $\theta$ measurement uncertainty of 5.2\degr, 
and $l\sim0.45$ pc, 
$B_\parallel \sim$200 $\mu$G is obtained.
Here we adopted $d=$260 pc, following \cite{gutermuth08} and \cite{bontemps10}.  
However, they also mentioned the possibility of a larger distance up to 700 pc.  
In case of the larger distance, $B$ becomes smaller by  
a factor of $(d/260 {\rm pc})^{-0.5}$ in these estimates.

The dynamical state of a magnetized cloud core is
measured by the ratio between the cloud mass and the 
magnetic flux, i.e., the mass-to-flux ratio, which 
is given by $M_{\rm cloud}/\Psi = \mu m_{\rm H}N/B$ 
$\sim$ 0.5--0.7 $\times (M_{\rm cloud}/\Psi)_{\rm critical}$
for these two zones, where $(M_{\rm cloud}/\Psi)_{\rm critical}$ 
is the critical value for a magnetic stability of the cloud 
 \citep[=$(4\pi^2 G)^{-1/2}$; ][]{nakano78}. 
Here, we assumed that the magnetic field is almost perpendicular to
the line-of-sight.
The estimated mass-to-flux ratios are close to the critical, implying that 
the magnetic field is likely to play an important role in the cloud dynamics 
and, thus,  star formation significantly.

\subsection{Conclusions}

We have presented near-infrared imaging polarimetry toward the Serpens South cloud.
The main findings are summarized as follows.

1. The $H$ and $K$s-band polarization measurements of near-infrared point sources 
indicated a well-ordered global magnetic field 
that is nearly perpendicular to the main filament with a mean position angle of $\sim50\degr$.
This imply that the magnetic field is likely to have controlled the formation of the main filament.  
On the other hand, the sub-filaments, which converge on the central part of the cluster, 
tend to run along the magnetic field, indicating a possibility that they are inflows or outflows 
along the magnetic field.

2. The global magnetic field appears to be curved in the southern part of the observed region. 
This curved morphology suggests an idea that the global magnetic field is distorted 
by gravitational contraction along the main filament toward the northern part where 
the mass of the Serpens South cloud seems to be mostly concentrated.

3. Applying the Chandrasekhar-Fermi method, the magnetic field strength is roughly estimated 
to be a few $\times$100 $\mu$G in two zones along the main filament.
The mass-to-flux rations in these zones indicate that the filamentary cloud is close to 
magnetically critical as a whole.

4. All the above results show that the magnetic field appears 
to significantly influence the dynamics of the Serpens South cloud, 
which is  associated with the cluster that is considered to be 
in the very early stage of cluster formation.  
This does not appear to support the weak magnetic field models of 
molecular cloud evolution/cluster formation \citep [e.g.,][]{padoan01}, 
at least not for the Serpens South cloud.

\acknowledgments

We are grateful for the support of the staff members of SAAO during the observation runs.  
K. S. thanks Y. Nakajima for assisting in the data reduction with the SIRPOL pipeline package.
This work was partly supported by Grant-in-Aid for Scientific Research (19204018, 
20403003) from the Ministry of Education, Culture, Sports, Science and Technology of Japan.

%%\clearpage

\end{document}